# *The role of symmetry in the interpretation of quantum mechanics*

Olimpia Lombardi and Sebastian Fortin

CONICET, Universidad de Buenos Aires

## 1. Introduction

After more than a century of its first formulations, quantum mechanics is still an arena of hot interpretive debates. However, despite the impressive amount of literature on the subject, the relevance of symmetry in interpretation is not properly acknowledged. In fact, although it is usually said that quantum mechanics is invariant under the Galilean group, the invariance is usually not discussed in the case of the interpretation of the theory. But this is a serious shortcoming in the foundational context, since the fact that a theory is covariant under a group does not guarantee the same property for interpretations to the extent that, in general, they add interpretive assumptions to the formal structure of the theory.

This situation may be considered irrelevant to those instrumentalist stances that are not interested in understanding what kind of world quantum mechanics describes. But it is urging for realist positions, which want to know how reality would be if quantum mechanics were true. From a realist perspective, it seems reasonable to require that an interpretation of quantum mechanics, to be admissible, preserves the same symmetries of the theory. In this paper we will take a realist view, in order to study what physical constraints the Galilean group imposes on interpretation. To this end, we will organize the presentation in two parts. In the first part we will consider the invariance properties of quantum mechanics: by beginning with considering the general notion of symmetry and the difference between the concepts of invariance and covariance, we will show that the Schrödinger equation is covariant under the Galilean group and we will consider under what conditions it is invariant under the same group. On this basis, in the second part we will address interpretive matters. First, we will consider the ontological meaning of invariance by stressing the relationship between invariance and objectivity. Then we will consider the constraints that Galilean invariance imposes onto any interpretation of quantum mechanics. These arguments will allow us, finally, to extrapolate the conclusions drawn for quantum mechanics to the case of quantum field theory.

## 2. Galilean symmetry of quantum mechanics

### 2.1 The general concept of symmetry

The idea of symmetry has a long history, during which it was associated with other notions such as harmony, equilibrium, beauty or proportion. At present, the everyday notion of symmetry is endowed with a geometric content that is familiar to everybody: something is symmetric when it has parts that are equal in a certain sense, such as in the case of the left-right symmetry of faces or the rotational symmetry of Escher's circle limit paintings.

The idea of symmetry acquires a precise definition in mathematics, when it is linked to the concept of invariance: from a mathematical viewpoint, an object is symmetric regarding a certain transformation when it is invariant under that transformation. But now, the transformation does not need to be geometric: the generic concept of symmetry applies to generic transformations in abstract mathematical spaces. The mathematical concept of symmetry was refined with the concept of group, which cluster different transformations into a specific structure. The concept of group was originally proposed by Galois in the first half of the nineteenth century, in the context of the resolution of algebraic equations by radicals. In the second half of he same century, Lie built a theory of continuous groups, with the purpose of extending Galois' methods for solving algebraic equations to differential equations. This work opened the way to apply the concepts of symmetry and invariance to the laws of physics expressed as mathematical equations.

Once the concept of symmetry is precisely defined in mathematical terms, different kinds of symmetry can be distinguished. One classification distinguishes between *global* and *local* symmetries: global symmetries depend on constant parameters, whereas local symmetries depend smoothly on the point of the base manifold. Another distinction is between *external or space-time* symmetries, and *internal or gauge* symmetries, due to invariance under non space-time transformations. The Galilean invariance of Newtonian mechanics and the Lorentz invariance of the special theory of relativity are paradigmatic examples of global space-time symmetries, which were called 'geometric' by Wigner (1967). Symmetries can also be classified as *continuous*, described by continuous or smooth transformations, or *discrete*, described by non-continuous transformations. Time-translation, space-translation and space-rotation are the traditional cases of continuous transformations, and time reversal, spatial reflection and charge conjugation are common examples of discrete transformations. Since in this paper we are interested in the Galilean group, we will only focus on global space-time continuous symmetries.

In principle, there are two possible interpretations of transformations: active and passive. Under the active interpretation, the transformation corresponds to a change from a system to another –transformed– system; for instance, one system translated in space with respect to the original one. Under the passive interpretation, the transformation consists in a change of the viewpoint –the reference frame– from which the system is described; for instance, the space-translation of the observer that describes the system. In the case of space-time transformations, continuous ones admit both interpretations, but active interpretation makes no sense in the case of discrete transformations (Sklar 1974: 363). Nevertheless, no matter the interpretation, the invariance of the fundamental law of a theory under its continuous symmetry group implies that the behavior of the system is not altered by the application of the transformation: in the active interpretation language, the original and the transformed systems are equivalent; in the passive interpretation language, the original and the transformed reference frames are equivalent.

## 2.2 Invariance and covariance

In the light of the general concept of symmetry, now the concept can be endowed with a more precise presentation. Although the link between symmetry and invariance is clear, it has not been explained yet to which items the property of invariance applies. As Brading and Castellani (2007) stress, the first step is to distinguish between symmetries of objects and symmetries of laws: "*we can apply the laws of mechanics to the evolution of our chair, considered as an isolated system, and these laws are rotationally invariant (they do not pick out a preferred orientation in space) even though the chair itself is not*" (Brading and Castellani 2007: 1332). In the case of physical laws, the symmetry of a law does not imply the symmetry of the objects (states and operators) contained in the equation that represents the law. Therefore, the conceptual implications of the symmetries of the law and of the involved objects under a particular group of transformations have to be both considered.

In the second place, it is necessary to consider the concept of covariance and its difference with invariance. In the literature on the subject there is no consensus about what 'covariance' means. Very commonly, the property of invariance is applied only to objects, and the property of covariance is reserved for laws and the corresponding equations. However, as Ohanian and Ruffini (1994) emphasize, the difference between the invariance and the covariance not only makes sense but also is relevant when applied to laws. In rough terms, an equation is invariant under a certain transformation when it does not change under the application of that transformation. In turn, an equation is covariant under a certain transformation when its form is left unchanged by that transformation (see Suppes 2000; Brading & Castellani, 2007). From an exclusively formal viewpoint, covariance is a rather weak property: any equation that is not covariant under a given transformation can always be expressed in a way such that makes it covariant by defining new functions of the original variables. However, covariance has physical significance only when those new functions can be endowed with physical meaning. In other words, if the transformation of the objects involved in a law is defined in advance due to physical reasons, one can decide univocally whether the law is invariant and/or covariant or not.

After these conceptual preliminaries, now we can introduce some formal definitions.

**Def. 1**: Let us consider a set $\mathcal{A}$ of objects $a_i \in \mathcal{A}$, and a group $G$ of transformations $g_\alpha \in G$, where the $g_\alpha : \mathcal{A} \to \mathcal{A}$ act on the $a_i$ as $a_i \to a'_i$. An *object* $a_i \in \mathcal{A}$ is *invariant under the transformation* $g_\alpha$ if, for that transformation, $a'_i = a_i$. In turn, the *object* $a_i \in \mathcal{A}$ is *invariant under the group* $G$ if it is invariant under all the transformations $g_\alpha \in G$.

In physics, the objects on which transformations apply are usually states $s$, observables $O$ and differential operators $D$, and each transformation acts on them in a particular way. Let us consider the example of time reversal on the objects involved in Hamilton equations: the state $s = (q,p)$, the observable Hamiltonian $H$, and the differential operators $D_1 = d/dt$, $D_2 = \partial/\partial p$ and $D_3 = \partial/\partial q$. The time-reversal transformation, which acts on the variable $t$ as $t \to -t$, reverses all the objects whose definitions in function of $t$ are non-invariant under the transformation:

$$s = (q,p) \to s' = (q',p') = (q,-p) \qquad O = H \to O' = H' \qquad D_1 = d/dt \to D'_1 = d'/dt = -d/dt$$

$$D_2 = \partial/\partial p \to D'_2 = \partial'/\partial p = -\partial/\partial p \qquad D_3 = \partial/\partial q \to D'_3 = \partial'/\partial q = \partial/\partial q$$

In physics, these objects on which transformations apply are combined in equations representing the laws of a theory. In particular, a dynamical law is represented by a differential equation $E(s,O_i,D_j) = 0$, which includes the state $s$, certain observables $O_i$ and certain differential operators $D_j$. When a transformation is applied to all these objects, the law may remain exactly the same, that is, its form may be left invariant by the transformation. This means that the relationship among the transformed objects is the same as that linking the original objects. But it may also be the case that the equation still holds when only the state is transformed, and this means that the evolution of the state is not affected by the transformation. Precisely:

**Def. 2**: Let $L$ be a *law* represented by an equation $E(s,O_i,D_j) = 0$, and let $G$ be a group of transformations $g_\alpha \in G$ acting on the objects involved in the equation as $s \to s'$, $O_i \to O'_i$ and $D_j \to D'_j$, $L$ is *covariant* under the transformation $g_\alpha$ if $E(s',O'_i,D'_j) = 0$, and $L$ is *invariant* under the transformation $g_\alpha$ if $E(s',O_i,D_j) = 0$. Moreover, $L$ is covariant –invariant– under the group $G$ if it is covariant –invariant– under all the transformations $g_\alpha \in G$.

On this basis, it is usually said that a certain group is the symmetry group of a theory:

> **Def. 3**: A group $G$ of transformations is said to be the *symmetry group* of a theory if the laws of the theory are covariant under the group $G$; this means that the laws preserve their validity even when the transformations of the group are applied to the involved objects.

Still in the case of the above example, the Hamilton equations, $d\boldsymbol{q}/dt = \partial H/\partial \boldsymbol{p}$ and $d\boldsymbol{p}/dt = -\partial H/\partial \boldsymbol{q}$, are covariant under time-reversal when $H' = H$, a condition satisfied when $H$ is time-independent; nevertheless, they are not invariant under time-reversal because $d\boldsymbol{p}'/dt \neq -\partial H/\partial \boldsymbol{q}$.

It is easy to see that, when a law is covariant under a transformation, and the observables and the differential operators contained in it are invariant under that transformation, the law is also invariant under the transformation. Nevertheless, as we will see in the particular case of the Schrödinger equation, this is not the only way to obtain the invariance of a law.

Some authors speak about symmetry instead of about covariance. For instance, Earman (2004a) defines symmetry in terms of the model of a theory.

> **Def. 4**: Let $\mathcal{M}$ be the set of the models of a certain mathematical structure, and let $\mathcal{M}_L \subset \mathcal{M}$ be the subset of the models satisfying the law $L$. A symmetry of the law $L$ is a map $S: \mathcal{M} \to \mathcal{M}$ that preserves $\mathcal{M}_L$, that is, for any $m \in \mathcal{M}_L$, $m' = S(m) \in \mathcal{M}_L$.

When $L$ is represented by a differential equation $E(s,O_i,D_j) = 0$, each model $m \in \mathcal{M}_L$ is represented by a solution $s = F(O_i,s_0)$ of the equation, corresponding to a possible evolution of the system. Then, the covariance of $L$ under a transformation $g$ –that is, the fact that $E(s',O'_i,D'_j) = 0$– implies that if $s = F(O_i,s_0)$ is a solution of the equation, $s' = F'(O'_i,s_0)$ is also a solution and, as a consequence, it represents a model $m' \in \mathcal{M}_L$. This means that the definition of covariance given by Def. 2 and the definition of symmetry given by Def. 4 are equivalent.

It is worth stressing that the covariance of a dynamical law –represented by a differential equation– does not imply the invariance of the possible evolutions –represented by the solutions of the equation–. Price (1996), illustrates this point in the case of time reversal with the familiar analogy of a factory which produces equal numbers of left-handed and right-handed corkscrews: the production as a whole is completely unbiased, but each individual corkscrew is spatially asymmetric (see Castagnino, Lara and Lombardi 2003, Earman 2004b). In fact, the covariance of the law $L$, represented by the equation $E(s, O_i, D_j) = 0$, implies that $s = F(O_i, s_0)$ and $s' = F'(O'_i, s_0)$ are both solutions of the equation, but does not imply that $s = s'$. In the model language, the symmetry of $L$ does not imply that $m = m'$. By contrast, invariance is a stronger property of the law: the invariance of $L$ means that $E(s', O_i, D_j) = 0$, in this case $s = s' = F(O_i, s_0)$ or, in the model language, $m = m'$.

The general definitions just introduced now allow us to explicitly state the conditions of covariance and invariance for quantum mechanics. Given a group $G$ whose transformations act on states, observables and differential operator as $|\varphi\rangle \to |\varphi'\rangle$, $|O\rangle \to |O'\rangle$ and $d/dt \to d'/dt$, the Schrödinger equation is *covariant* when

$$\frac{d'|\varphi'\rangle}{dt} = -i\hbar H' |\varphi'\rangle \tag{1}$$

and it is *invariant* when

$$\frac{d|\varphi'\rangle}{dt} = -i\hbar H |\varphi'\rangle \tag{2}$$

## 2.3 The Galilean group

As emphasized by Lévi-Leblond (1974), although the covariance –and even the invariance– of non-relativistic quantum mechanics under the Galilean transformations is usually assumed as a well-known fact, in general this conceptual issue is absent from the standard literature about the theory: only in very few cases this assumption is grounded on a conceptual elucidation of the involved notions. With the exception of the book of Ballentine (1998), it is common to see that the Galilean group is not even mentioned in the textbooks on the matter. For this reason, it is worth dwelling on this point.

Under the assumption that time can be represented by a variable $t \in \mathbb{R}$ and position can be represented by a variable $\boldsymbol{r} = (x, y, z) \in \mathbb{R}^3$, the Galilean group $\mathcal{G} = \{g_\alpha\}$, with $\alpha = 1$ to $10$, is a group of continuous space-time transformations $g_\alpha : \mathbb{R}^3 \times \mathbb{R} \to \mathbb{R}^3 \times \mathbb{R}$ acting as

- ➢ Time-translation: $\quad t \to t' = t + \tau$
- ➢ Space- translation: $\quad \boldsymbol{r} \to \boldsymbol{r}' = \boldsymbol{r} + \boldsymbol{\rho}$
- ➢ Space-rotation: $\quad \boldsymbol{r} \to \boldsymbol{r}' = R_\theta \boldsymbol{r}$
- ➢ Velocity-boost: $\quad \boldsymbol{r} \to \boldsymbol{r}' = \boldsymbol{r} + \boldsymbol{u} t$

where $\tau \in \mathbb{R}$ is a real number representing a time interval, $\boldsymbol{\rho} = (\rho_x, \rho_y, \rho_z) \in \mathbb{R}^3$ is a triple of real numbers representing a space interval, $R_\theta \in \mathcal{M}^{3\times 3}$ is a $3\times 3$ matrix representing a space-rotation an angle $\theta$, and $\boldsymbol{u} = (u_x, u_y, u_z) \in \mathbb{R}^3$ is a triple of real numbers representing a constant velocity.

Since the Galilean group $\mathcal{G}$ is a Lie group, the Galilean transformations $g_\alpha$ can be represented by unitary operators $U_\alpha$ over the Hilbert space, with the exponential parametrization $U_\alpha = e^{iK_\alpha s_\alpha}$, where $s_\alpha$ is a continuous parameter and $K_\alpha$ is a Hermitian operator independent of $s_\alpha$, called *generator* of the transformation $g_\alpha$. Therefore, the Galilean group $\mathcal{G}$ is defined by ten group generators $K_\alpha$: one time-translation $K_\tau$, three space-translations $K_{\rho_i}$, three space-rotations $K_{\theta_i}$, and three velocity-boosts $K_{u_i}$, with $i = x, y, z$. The generators of $\mathcal{G}$ form the Galilean algebra, that is, the Lie algebra of the Galilean generators. The combined action of all the transformations is given by

$$U_s = \prod_{\alpha=1}^{10} e^{iK_\alpha s_\alpha} \tag{3}$$

Strictly speaking, in the case of quantum mechanics the symmetry group is the group corresponding to the central extension of the Galilean algebra, obtained as a semi-direct product between the Galilean algebra and the algebra generated by a central charge, which in this case denotes the mass operator $M = mI$, where $I$ is the identity operator and $m$ is the mass. The mass operator as a central charge is a consequence of the projective representation of the Galilean group (see Weinberg 1995, Bose 1995). However, in order to simplify the presentation, from now on we will use the expression 'Galilean group' to refer to the corresponding central extension, and we will take $\hbar = 1$ as usual.

Since the Galilean group is a Lie group, it is defined by the commutation relations between its generators:

(a) $\left[K_{\rho_i}, K_{\rho_j}\right] = 0$      (f) $\left[K_{u_i}, K_{\rho_j}\right] = i\delta_{ij} M$

(b) $\left[K_{u_i}, K_{u_j}\right] = 0$      (g) $\left[K_{\rho_i}, K_\tau\right] = 0$

(c) $\left[K_{\theta_i}, K_{\theta_j}\right] = i\varepsilon_{ijk} K_{\theta_j}$      (h) $\left[K_{\theta_i}, K_\tau\right] = 0$

(d) $\left[K_{\theta_i}, K_{\rho_j}\right] = i\varepsilon_{ijk} K_{\rho_k}$      (i) $\left[K_{u_i}, K_\tau\right] = iK_{\rho_i}$

(e) $\left[K_{\theta_i}, K_{u_j}\right] = i\varepsilon_{ijk} K_{u_k}$ (4)

where $\varepsilon_{ijk}$ is the Levi-Civita tensor, such that $i \neq k$, $j \neq k$, $\varepsilon_{ijk} = \varepsilon_{jki} = \varepsilon_{kij} = 1$, $\varepsilon_{ikj} = \varepsilon_{jik} = \varepsilon_{kji} = -1$, and $\varepsilon_{ijk} = 0$ if $i = j$. In quantum mechanics, when the system is free from external fields, the generators $K_\alpha$ represent the basic magnitudes of the theory: the energy $H = \hbar K_\tau$, the three momentum components $P_i = \hbar K_{\rho_i}$, the three angular momentum components $J_i = \hbar K_{\theta_i}$, and the three boost components $G_i = \hbar K_{u_i}$. Then, by taking $\hbar = 1$, the commutation relations result

(a) $[P_i, P_j] = 0$     (f) $[G_i, P_j] = i\delta_{ij} M$
(b) $[G_i, G_j] = 0$     (g) $[P_i, H] = 0$
(c) $[J_i, J_j] = i\varepsilon_{ijk} J_k$     (h) $[J_i, H] = 0$
(d) $[J_i, P_j] = i\varepsilon_{ijk} P_k$     (i) $[G_i, H] = iP_i$
(e) $[J_i, G_j] = i\varepsilon_{ijk} G_k$     (5)

The rest of the physical magnitudes can be defined in terms of these basic ones: for instance, the three position components are $Q_i = G_i / m$, the three orbital angular momentum components are $L_i = \varepsilon_{ijk} Q_j P_k$, and the three spin components are $S_i = J_i - L_i$.

In the Hilbert formulation of quantum mechanics, each Galilean transformation $g_\alpha \in \mathcal{G}$ acts on states and on observables as

$$|\varphi\rangle \to |\varphi'\rangle = U_{s_\alpha} |\varphi\rangle = e^{iK_\alpha s_\alpha} |\varphi\rangle \tag{6}$$

$$O \to O' = U_{s_\alpha} O U_{s_\alpha}^{-1} = e^{iK_\alpha s_\alpha} O e^{-iK_\alpha s_\alpha} \tag{7}$$

In turn, the invariance of an observable $O$ under a Galilean transformation $g_\alpha$ amounts to the commutation between $O$ and the corresponding generator $K_\alpha$:

$$O' = e^{iK_\alpha s_\alpha} O e^{-iK_\alpha s_\alpha} = O \iff [O, K_\alpha] = 0 \tag{8}$$

## 2.4 Invariance and covariance in quantum mechanics

In order to decide about the Galilean covariance and invariance of quantum mechanics, it is necessary to analyze how the Galilean transformations act on the Schrödinger equation. In fact, the action of a generic $U = e^{iKs}$ results in

$$\frac{d|\varphi'\rangle}{dt} = -i\left[H' + i\frac{dU}{dt} U^{-1}\right] |\varphi'\rangle \tag{9}$$

*a) The invariance of the Schrödinger equation.*
In a closed, constant-energy system free from external fields, $H$ is time-independent and the $P_i$ and the $J_i$ are constants of motion (see eqs. (5g,h)). Then, for time-translations, space-translations and space-rotations, $dU/dt = de^{iKs}/dt = 0$, where $K$ and $s$ stand for $H$ and $\tau$, $P_i$ and $\rho_i$, and $J_i$ and $\theta_i$, respectively. As a consequence, eq. (9) yields

$$\frac{d|\varphi'\rangle}{dt} = -iH'|\varphi'\rangle \tag{10}$$

Moreover, since in this closed-system case $H$ commutes with $P_i$ and $J_i$ (see eqs. (5g,h)), for those three transformations $H' = H$ (see eq. (8)). By applying this result to eq. (10), we obtain eq. (2) and, so, we prove the invariance of the Schrödinger equation under time-translations, space-translations and space-rotations when there are no external fields acting on the system.

The case of boost-transformations is different from the previous cases, because the Hamiltonian is not boost-invariant even when the system is free from external fields (for the same claim in classical mechanics, see Butterfield, 2007, p. 6). In fact, under a boost-transformation corresponding to a velocity $u_x$, since $[G_x, H] = iP_x \neq 0$ (eq. (5i)), $H$ changes as

$$H' = e^{iG_x u_x} H e^{-iG_x u_x} \neq H \tag{11}$$

and the generator $G_x$ is

$$G_x = mQ_x = m(Q_{x0} + V_x t) = mQ_{x0} + P_x t \tag{12}$$

Since $G_x$ is not time-independent, $dU/dt = de^{iG_x u_x}/dt \neq 0$, and eq. (4-10) yields

$$\frac{d|\varphi'\rangle}{dt} = -i\left[H' + i\frac{de^{iG_x u_x}}{dt}e^{-iG_x u_x}\right]|\varphi'\rangle \tag{13}$$

When the value of the bracket in the r.h.s. side of eq. (13) is computed, it can be proved that the terms added to $H$ in $H'$ cancel with those coming from the term containing the time-derivative (see Lombardi, Castagnino and Ardenghi 2010). Therefore, eq. (2) is again obtained and the invariance of the Schrödinger equation is proved also for boost-transformations.

In summary, when there are no external fields acting on the system, the Hamiltonian is invariant under time- translations, space- translations and space-rotations, but not under boost-transformations.

When the system is under the action of external fields, the fields modify the evolution of the system: for example, in the case of a non-isotropic potential, it cannot longer be expected that the system does not modify its behavior when rotated in space. But, in non-relativistic quantum mechanics, fields are not quantized: they do not play the role of quantum systems that interact with other systems. For this reason, the effect of the fields on a system has to be included in its Hamiltonian, because it is the only observable involved in the time-evolution law. As a consequence, under the action of fields the Hamiltonian is no longer the generator of time- translations: it only retains its role of generator of the dynamical evolution (see Laue 1996, Ballentine 1998). Therefore, the commutation relations involving the Hamiltonian, eqs. (5g,h,i), are no longer valid: now these relations hold with the generator of time-translations $d/dt$ (see eqs. (4g,h,i)), but not with the Hamiltonian. Therefore, the time-independence of the $P_i$ and the $J_i$ cannot be guaranteed. As a consequence, in the general case, the Schrödinger equation is not Galilean invariant in the presence of external fields.

*b) The covariance of the Schrödinger equation.*

In order to study the covariance of the Schrödinger equation, let us rewrite eq. (9) as

$$\frac{d|\varphi'\rangle}{dt} - \frac{dU}{dt}U^{-1}|\varphi'\rangle = -iH'|\varphi'\rangle \tag{14}$$

This shows that the equation is covariant because the differential operator transforms as

$$\frac{d}{dt} \to \frac{d'}{dt} = \frac{d}{dt} - \frac{dU}{dt}U^{-1} = \frac{D}{Dt} \tag{15}$$

In other words, the transformed differential operator $d'/dt$ is a covariant time-derivative $D/Dt$, which makes the Schrödinger equation to be Galilean-covariant in the following sense (see eq. (1))

$$\frac{d'|\varphi'\rangle}{dt} = \frac{D|\varphi'\rangle}{Dt} = -iH'|\varphi'\rangle \tag{16}$$

As shown above, without external fields, $H$, the $P_i$ and the $J_i$ are time-independent and, as a consequence, $dU/dt = 0$; then, eq. (15) shows that $d/dt$ is invariant under time-translations, space-translations and space-rotations. But since for boost-transformations this is not the case, the covariance of the Schrödinger equation requires the transformation of the differential operator as $d/dt \to D/Dt$: covariance under boosts amounts to a sort of "non-homogeneity" of time that requires the covariant adjustment of the time-derivative. This illustrates the claim advanced in Subsection 2.2: although a law is invariant under a transformation when it is covariant and all the involved objects are invariant, this is not the only way to obtain invariance. When the system is free from external fields, the Schrödinger equation is invariant under boost-transformations, in spite of the fact that the Hamiltonian and the differential operator $d/dt$ are not boost-invariant objects.

When external fields are applied on the system, the Hamiltonian includes the action of the fields. Then, although eq. (16) is still valid, the transformed Hamiltonian $H' = UHU^{-1}$ has to be computed case by case, and the conditions that the external potentials have to satisfy in order to preserve covariance can be deduced (see Brown and Holland 1999, Colussi and Wickramasekara 2008).

# 3. Invariance in interpretation

## 3.1 The ontological meaning of symmetry

As it is usually accepted, the Galilean group is the symmetry group of continuous space-time transformations of classical and quantum mechanics. In the language of the passive interpretation, the invariance of the dynamical laws amounts to the equivalence among inertial reference frames (time-translated, space-translated, space-rotated or uniformly moving with respect to each other). In other words, Galilean transformations do not introduce a modification in the physical situation, but only express *a change in the perspective from which the system is described*.

This merely perspectival meaning of the Galilean symmetries depends, in turn, on the properties of space and time. Invariance under time-displacements expresses the homogeneity of time; invariance under space-displacements and/or space-rotations expresses the homogeneity and/or the isotropy of space, respectively. These invariances are embodied in the commutation relations that define the Galilean group (see eqs. (5)). Nevertheless, space is not always homogeneous and isotropic. In non-relativistic quantum mechanics, fields are not quantized: they are treated as classical fields that act on the quantum system by breaking the homogeneity and/or the isotropy of space. This breaking turns out to be expressed in the form of the Hamiltonian: the non-homogeneity of space leads to the fact that, at least, some $P_i$ is not a constant of motion ($[P_i, H] \neq 0$); the non-isotropy of space leads to the fact that, at least, some $J_i$ is not a constant of motion ($[J_i, H] \neq 0$). And this, in turn, amounts to the breaking of the full validity of the Galilean group under the form of eqs. (5): the commutation relations involving the Hamiltonian ((5g), (5h) and (5i)) are, in general, no longer valid. In this case, the commutation relations are still defined by eqs. (4), but the generators of space-translations and space-rotations are not $P$ and $J$, but have to be defined in each case, depending on the specific form of the external field.

The above remarks are related with the fact that certain quantities are *physically irrelevant* in the light of the symmetries of a theory. For instance, the space-translation symmetry of a dynamical law means that where the system is particularly located in space is irrelevant to its evolution governed by that law. The notion of physical irrelevance endows with physical content the difference between local and global symmetries: "*A global symmetry reflects the irrelevance of absolute values of a certain quantity: only relative values are relevant*" (see Brading and Castellani 2007: 1360). For instance, in classical mechanics, for example, space-translation invariance implies that absolute position is irrelevant to the system's behavior: the equations of motion do not depend on absolute positions, only relative positions matter. The physical irrelevance of certain quantities is strongly linked with the issue of *objectivity*.

The intuition about a strong link between invariance and objectivity is rooted in a natural idea: what is objective should not depend on the particular perspective used for the description. When this intuition is translated to group-theoretical language, it can be said that what is objective according to a theory is what is invariant under the symmetry group of the theory. This idea is not new. In the domain of formal sciences, already Felix Klein, in his "Erlangen Program" of 1872, tried to characterize all known geometries by their invariants, that is, by the quantities which are not changed under a particular group of transformations (see Kramer 1970). This idea passed to physics with the advent of relativity: it was widely discussed in the context of special and general relativity with respect to the ontological status of space and time. In Minkowski words: "*Henceforth space for itself, and time by itself, are doomed to fade away into mere shadows, and only a kind of union of the two will preserve an independent reality*" (Minkowski, 1923, p. 75). The claim that objectivity means invariance begins to appear in Weyl's works, applied to mathematics, in his *Philosophy of Mathematics and Natural Science* (1927), when he claims that "*A point relation is called objective if it is invariant under all automorphisms*" (cited in Vollmer 2010: 1661). The idea, applied to physical sciences, becomes a main thesis of his book *Symmetry* (Weyl 1952). In recent times, the idea has strongly reappeared in several works. For instance, in her deep analysis of quantum field theory, Auyang (1995) makes her general concept of "object" to be founded on its invariance under transformations among all representations. In turn, the assumption that invariance is the root of objectivity is the central theme of Nozick's book *Invariances: The Structure of the Objective World* (2001).

Once the ontological meaning of symmetry is acknowledged, it is easy to admit that symmetry must play a relevant role in the understanding of a physical theory. In the particular case of quantum mechanics, once it is seen in what sense the Galilean group is the symmetry group of the theory, the consideration of Galilean invariance cannot be overlooked in the discussions about interpretation.

**3.2 An invariant interpretation of quantum mechanics**

The physical meaning of the action of the Galilean transformations is well-understood in classical mechanics. However, as pointed out in the Introduction, this issue is scarcely discussed in the field of quantum mechanics, perhaps under the assumption that the matter is as easy as in the classical case. But we will see that quantum mechanics is peculiar also regarding to this point.

As it is well known, Heisenberg's uncertainty principle poses a fundamental limit to the precision with which certain pairs of physical observables –complementary observables– can be known simultaneously. Nevertheless, this result leaves open the way to think in the possibility of completing the theory with certain "hidden variables", which would determine the values of all the observables of the system at any time, in a classical-like manner. The Kochen-Specker theorem (1967) breaks this possibility down by putting a barrier to any realist classical-like interpretation of quantum mechanics. In fact, the theorem proves the impossibility of ascribing precise values to *all* the physical quantities (observables) of a quantum system simultaneously, while preserving the functional relations between commuting observables. In other words, this result is a manifestation of the *contextuality* of quantum mechanics: the ascription of precise values to the observables of a quantum system is always contextual.

As a consequence of the Kochen-Specker theorem, any realist interpretation of quantum mechanics is committed to selecting a "privileged" set of observables out of all the observables of the system. The observables of that set will be those that acquire a definite value without breaking quantum contextuality. At this point, the symmetry group of the theory enters the scene: as stressed by Brown, Suárez and Bacciagaluppi (1998), any interpretation that selects the set of the definite-valued observables of a quantum system in a given state is committed to considering how that set is transformed under the Galilean group.

But now the link between invariance and objectivity comes into play. The study of the role of symmetry is particularly urging in the case of realist interpretations of quantum mechanics, which conceive a definite-valued observable as a physical magnitude that objectively acquires an actual value among all its possible values: the fact that a certain observable acquires a definite value has to be an objective fact. Therefore, since the invariance of the theory holds, the set of the definite-valued observables of a system picked out by the interpretation should be left invariant by the Galilean transformations: from a realist viewpoint, it would be unacceptable that such a set changed as the mere result of a change in the perspective from which the system is described.

The natural way to reach this goal is to appeal to the *Casimir operators of the Galilean group*: if the interpretation has to select a Galilean-invariant set of definite-valued observables, such a set must depend on those Casimir operators, since they are invariant under all the transformations of the Galilean group. The –central extension of the– Galilean group has three Casimir operators which, as such, commute with all the generators of the group: they are the mass operator $M$, the operator $S^2$, and the internal energy operator $W = H - P^2/2m$. The eigenvalues of the Casimir operators label the irreducible representations of the group; so, in each irreducible representation, the Casimir operators are multiples of the identity: $M = mI$, $S^2 = s(s+1)I$, where $s$ is the eigenvalue of the spin $S$, and $W = wI$, where $w$ is the scalar internal energy.

Whereas the fact that the system objectively acquires a definite value of the mass and the spin seems strongly reasonable, the fact that the Hamiltonian is not included in the "privileged" set may sound puzzling, given the very special role that the Hamiltonian plays in quantum mechanics by ruling the time-evolution of quantum systems. So, it is worth taking a while to consider how the Hamiltonian behaves under the action of the Galilean transformations.

Let us consider a quantum system not affected by external fields, whose Hamiltonian, in a generic reference frame $RF$, reads $H = P^2/2m + W = K + W$, where the kinetic energy $K = P^2/2m$ only depends on the total momentum relative to $RF$, and the internal energy $W$ does not depend on the position and the momentum relative to $RF$, but only depends on differences of positions and, eventually, on their derivatives. It is precisely these features of $K$ and $W$ what guarantees that $[K,W] = 0$ and, as a consequence, $H$ can be expressed as

$$H = P^2/2m + W = K + W = H_K \otimes I_W + I_K \otimes H_W \qquad (17)$$

where $H_K$ is the kinetic Hamiltonian acting on a Hilbert space $\mathcal{H}_K$, $H_W$ is the internal energy Hamiltonian acting on a Hilbert space $\mathcal{H}_W$, and $I_K$ and $I_W$ are the identity operators of the respective tensor-product spaces (for examples in well-known models, see Ardenghi, Castagnino & Lombardi, 2009). As stressed above, the Hamiltonian is invariant under time-displacements, space-displacements and space-rotations, but not under boost-transformations; so let us consider that case.

If a boost-transformation of velocity $u_x$ is applied to the system, the unitarily transformed Hamiltonian is (see proof in Lombardi, Castagnino and Ardenghi 2010)

$$H' = e^{iG_x u_x} H e^{-iG_x u_x} = H - u_x P_x + \frac{1}{2} M u_x^2 = H + T_B \qquad (18)$$

where $T_B$ is the boost contribution to the energy. Therefore, it can be expressed as

$$H' = H + T_B = \frac{P^2}{2m} + W + T_B = K' + W \qquad (19)$$

where $K'$ is the transformed kinetic energy (see eqs. (4-18) and (4-20)):

$$K = \frac{P^2}{2m} \quad \Rightarrow \quad K' = K + T_B = \frac{P^2}{2m} + T_B = \frac{(P + P_B)^2}{2m} \qquad (20)$$

For the same reasons as before, $[K',W]=0$ and, as a consequence, $H'$ can be written as

$$H' = K' + W = H'_K \otimes I_W + I_K \otimes H_W \tag{21}$$

where $H'_K = H_K + H_B$ is the transformed kinetic Hamiltonian acting on $\mathcal{H}_K$. In other words,

$$H'_W = H_W \qquad H'_K = H_K + H_B \tag{22}$$

This means that the application of a boost-transformation does not modify the internal energy $W$ of the system: $W$ is boost-invariant, in agreement with the fact that it is a Casimir operator of the Galilean group and that it only depends on differences of positions (it is a "relevant" quantity). The boost-transformation only modifies the kinetic energy by adding the kinetic energy of the boost, in agreement with the fact that it is not a Casimir operator of the Galilean group and that it depends on the velocity relative to the reference frame $RF$ (it is an "irrelevant" quantity).

The above considerations all point to the same direction: the *objective* content of the energy description of the system is given by the internal energy $W$, which is invariant under the whole Galilean group. On the contrary, the kinetic energy $K$, whose value is modified by a boost, can be viewed as a non-objective magnitude that changes with the mere change of the descriptive perspective. In particular, when the system is described in the reference frame at rest with respect to its center of mass, then $P = 0$ and the kinetic energy disappear from the description.

These conclusions about the non-objectivity of the kinetic energy are not challenged by the fact that a boost-transformation has well-defined manifestation in the energy spectrum of the system, since it produces a Doppler shift on that spectrum. However, we also know that energy is defined up to a constant value: the relevant information about the energy spectrum of a system is contained in its internal energy, and the kinetic energy only introduces a shift of that spectrum. Therefore, the internal energy carries the physically meaningful structure of the energy spectrum, and the kinetic energy represents an energy shift which, although observable, is physically non relevant and merely relative to the reference frame used for the description.

Recently, a new interpretation of quantum mechanics has exploited the symmetry features of the theory to solve its main conceptual conundrums. The modal-Hamiltonian interpretation (Lombardi and Castagnino 2008, Castagnino and Lombardi 2008, Lombardi 2010, Ardenghi and Lombardi 2011, Lombardi, Fortin, Castagnino and Ardenghi 2012) is a realist, non-collapse approach according to which the quantum state describes the possible properties of the system but not its actual properties. According to this interpretation, the Hamiltonian is decisive in the definition of systems and subsystems, and in the selection of the preferred context where observables acquire definite values. This interpretation has been applied to many well-known physical situations (free particle, free particle with spin, harmonic oscillator, free hydrogen atom, Zeeman effect, fine structure, the Born-Oppenheimer approximation), leading to results consistent with experimental evidence (Lombardi and Castagnino 2008, Section 5). Moreover, it has proved to be effective for solving the measurement problem, both in its ideal and its non-ideal versions. (Lombardi and Castagnino 2008, Section 6, Ardenghi, Lombardi and Narvaja 2013, Lombardi, Fortin and López 2015). This interpretive view also promotes an ontology of properties, based on the algebraic approach to QM, where systems are bundles of properties represented by quantum observables (da Costa, Lombardi and Lastiri 2013, da Costa and Lombardi 2014, Lombardi and Dieks 2016).

Although based on the central role of the Hamiltonian, the modal-Hamiltonian interpretation was reformulated in an explicitly invariant form, according to which the definite-valued observables of a quantum system free from external fields are the observables $C_i$ represented by the Casimir operators of the Galilean group in the corresponding irreducible representation, and all the observables commuting with the $C_i$ and having, at least, the same symmetries as the $C_i$ (Lombardi, Castagnino and Ardenghi 2010). In turn, as argued above, from a realist viewpoint, the fact that certain observables acquire an actual definite value is an objective fact in the behavior of the system; therefore, the set of definite-valued observables selected by a realist interpretation must be also Galilean-invariant. But the Galilean-invariant observables are always functions of the Casimir operators of the Galilean group. As a consequence, one is led to the conclusion that any realist interpretation that intends to preserve the objectivity of the set of the definite-valued observables may not stand very far from the modal-Hamiltonian interpretation.

### 3.3 Invariance and interpretation in quantum physics

In his paper "Physical reality," Born (1953) expressed very clearly his conviction about the strong link between invariance and objectivity: "*I think the idea of invariance is the clue to a rational concept of reality*" (1953: 144); "*The main invariants are called charge, mass (or rather: rest-mass), spin, etc.; and in every instance, when we are able to determine these quantities, we decide to have to do with a definite particle. I maintain that we are justified in regarding these particles as real in a sense not essentially different from the usual meaning of the word.*" (1953: 146).

Born's words suggest us the possibility of generalize the idea developed in this work in two senses. In non-relativistic quantum mechanics, the external fields acting on a system are not quantized, and this fact is what breaks down the harmony of the free case: the Hamiltonian is no longer the generator of time- translations in the Galilean group. In quantum field theory, on the contrary, fields are quantum items and not "external" fields affecting the behavior of the quantum system. As a consequence, the generators of the Poincaré group do not need to be reinterpreted in the presence of "external" factors. These features of quantum field theory make us to consider whether the realist interpretation, expressed in terms of the Casimir operators of the Galilean group in non-relativistic quantum mechanics, can be transferred to quantum field theory by changing accordingly the symmetry group: the definite-valued observables of a system in quantum field theory would be those represented by the Casimir operators of the Poincaré group. Since $M$ and $S^2$ are the only Casimir operators of the Poincaré group, they would always be definite-valued observables. This conclusion would stand in agreement with a usual physical assumption in quantum field theory: elemental particles always have definite values of mass and spin, and those values are precisely what define the different kinds of elemental particles of the theory. Moreover, the classical limit of quantum field theory manifests the limit of the corresponding Casimir operators (see Ardenghi, Castagnino and Lombardi 2011): there is a meaningful limiting relation between the observables that acquire definite values according to quantum field theory and those that acquire definite values according to quantum mechanics.

But the idea can also be generalized in a second sense: if invariance is a mark of objectivity, it should guide the interpretation not only of quantum mechanics, but also of any physical theory with definite symmetries. Following this idea, there is no reason to focus only on space-time global symmetries: internal or gauge symmetries should also be considered as relevant in the definition of objectivity and, as a consequence, in the identification of the definite-valued observables of the system. For instance, in relativistic quantum mechanics a gauge symmetry is what identifies the charge as an objective quantity. Therefore, the generalized principle for interpreting quantum theories from a realistic viewpoint can be stated as follows: the definite-valued observables of a system whose behavior is governed by a certain quantum theory are the observables invariant under all the transformations corresponding to the symmetries of the theory, both external and internal.

## 4. Conclusions

In this paper we focused on a question usually not taken into account in the literature on the interpretation of quantum mechanics in particular and quantum physics in general: the question about how an interpretation should behave under the symmetry group of the theory. By echoing the widespread position that links invariance and objectivity, and by considering that, from a realist viewpoint, it is unacceptable that what acquire definite value changes as the mere result of a change in the perspective from which the system is described, we have proposed a definite interpretive principle: the objective definite-valued observables of a quantum system are the observables invariant under all the transformations corresponding to the symmetries of the theory that governs its behavior. We have introduced a particular interpretation of quantum mechanics that satisfies this general principle. Nevertheless, the proposal of this work goes beyond a particular interpretation, since it intends to supply a general framework that guides the building of any realist interpretation in the light of the physically central concept of symmetry.